\documentclass[showpacs]{revtex4}
\usepackage{subfigure}
\usepackage{graphicx}

\begin{document}

\title{Mismatch management for optical and matter-wave quadratic solitons}
\author{R. Driben, Y. Oz, B. A. Malomed, and A. Gubeskys}
\affiliation{Department of Interdisciplinary Studies, School of Electrical Engineering,
Faculty of Engineering, Tel Aviv University, Tel Aviv 69978, Israel}
\author{V. A. Yurovsky }
\affiliation{School of Chemistry, Faculty of Exact Sciences, Tel Aviv University, Tel
Aviv 69978, Israel}

\begin{abstract}
We propose a way to control solitons in $\chi ^{(2)}$
(quadratically-nonlinear) systems by means of periodic modulation
imposed on the phase-mismatch parameter (``mismatch management",
MM). It may be realized in the co-transmission of
fundamental-frequency (FF) and second-harmonic (SH) waves in a
planar optical waveguide via a long-period modulation of the usual
quasi-phase-matching pattern of ferroelectric domains. The MM may
also be implemented by dint of the Feshbach resonance in a
harmonically-modulated magnetic field in a hybrid atomic-molecular
Bose-Einstein condensate (BEC), with the atomic and molecular mean
fields (MFs) playing the roles of the FF and SH, respectively. The
problem is analyzed by two methods. First, we identify stability
regions for spatial solitons in the MM system, in terms of the MM
amplitude and period, using the MF equations for
spatially-inhomogeneous configurations. In particular, an
instability enclave is found inside the stability area.The
robustness of the solitons is also tested against variation of the
shape of the input pulse, and a threshold for the formation of
stable solitons is found in terms of its power. Interactions
between solitons are virtually unaffected by the MM. The second
method (\textit{parametric approximation}), going beyond the MF
description, is developed for spatially-homogeneous states. It
demonstrates that the MF description is valid for large modulation
periods, while at smaller periods the non-MF component acquires
gain, which implies destruction of MF under the action of the
high-frequency MM.
\end{abstract}

\maketitle

\section{Introduction}

Solitons are robust localized pulses that have been predicted
theoretically and created experimentally in diverse physical
settings. The current research in this field is heavily focused on
nonlinear optics \cite{KA} and Bose-Einstein condensation (BEC)
\cite{ReviewRandy}. Typical solitons are found in uniform media
with constant characteristics. However, in many cases it is
necessary to consider solitary waves (which are also called
``solitons", in a loose sense, even if they are not described by
integrable equations) traveling across heterogeneous media, or
subjected to strong time modulation. A well-known example of the
former setting is \textit{dispersion management}, which is an
important concept in fiber-optic telecommunications, helping to
support stable soliton trains used as data-carrying streams
\cite{DM,book}. On the other hand, a
possibility to stabilize matter-wave solitons by means of the \textit{%
nonlinearity management}, applied to them via the Feshbach resonance in a
modulated magnetic field (i.e., time-periodic variation of the scattering
length that determines the coefficient in front of the cubic term in the
Gross-Pitaevskii equation), in one-dimensional (1D) \cite{FRM} and 2D \cite%
{Fatkhulla} geometries, has drawn considerable attention in the studies of
BEC. Another example of a setting supporting the transmission of robust
solitons in a strongly heterogeneous periodic system, that combines features
of both the dispersion management and nonlinearity management, is the \emph{%
split-step model} (SSM). In the simplest case, it is composed of
periodically alternating pieces of optical fibers with zero dispersion and
zero nonlinearity (i.e., it is built as a periodic concatenation of
nonlinear and dispersive segments, the latter ones taken with anomalous
dispersion) \cite{SSM,book}. In a more realistic variant of the SSM, the
nonlinear and dispersive segments are allowed to have nonvanishing
dispersion and nonlinearity, respectively \cite{SSM2}. Multicomponent
generalizations of the SSM were elaborated too, including one for the WDM
(wavelength-division-multiplexed) system \cite{WDM}, and a model taking into
regard two polarizations of light \cite{PMD}.

In the above-mentioned examples, solitons are supported by the cubic
nonlinearity of the medium. It is well known that the quadratic
(second-harmonic-generating, alias $\chi ^{(2)}$) nonlinearity also gives
rise to stable solitons, that have been studied in detail theoretically and
experimentally in optics \cite{chi2}. In most cases, these are \textit{%
spatial solitons}, i.e., self-supporting localized beams in bulk or (which
is more relevant to the present work) planar waveguides. Temporal $\chi
^{(2)}$ solitons have been created too \cite{Paolo,chi2}, but under very
sophisticated conditions.

In terms of BEC, a counterpart of the second-harmonic generation is the
Feshbach association in atomic BEC \cite{TTHK99,StoofReview,KGJ06}, induced
by the coupling of atomic and molecular mean fields (MFs) with the help of
resonant optical fields or by hyperfine interactions, using the Zeeman
effect for the mismatch tuning. Quadratic solitons in BEC have been
predicted in Ref.\ \cite{Peter}. In terms of the comparison with optics,
these solitons in BEC may be classified as temporal ones.

In those studies, it was established that conditions for the
existence and stability of $\chi ^{(2)}$ solitons are most
sensitive to the mismatch between the fundamental-frequency (FF)
wave and the second harmonic (SH), or, in terms of the BEC,
between the atomic and molecular MFs. This fact suggests a natural
question, which is the subject of the present paper -- how the
quadratic solitons will react to a periodic modulation of the
mismatch, i.e., ``mismatch management" (MM). The issue is of
general interest, as a possible contribution to the theory of the
``soliton management" \cite{book}, and may also be potentially
promising as concerns the use of $\chi ^{(2)}$ spatial solitons in
optical switching \cite{chi2} and other applications to photonics.
As for the soliton-management schemes, almost all of them were
explored in terms of media with the cubic nonlinearity; the only
example dealing with a $\chi ^{(2)}$ setting was the model of
``tandem solitons", which assumed their transmission in a
waveguide built as a concatenation of $\chi ^{(2)}$and linear
segments, in 1D \cite{Lluis} and 2D \cite{Dumitru} geometry
(actually, the purpose of the tandem model was to reduce the
mismatch).

A ubiquitous approach to the reduction of mismatch in optical waveguides is\
based on the use of the quasi-phase-matching (QPM)\ scheme, in which the
material of the $\chi ^{(2)}$ waveguide is subjected to periodic poling
(since the material, such as LiNbO$_{3}$, is a ferroelectric, this is
usually carried out through periodic reversal of the orientation of
ferroelectric domains) \cite{QPM}. The poling gives rise to a change of the
sign of the $\chi ^{(2)}$ coefficient with a certain period, $L_{\mathrm{QPM}%
}$, and thus adds an extra wave vector, $\mathbf{k}_{\mathrm{QPM}%
}=\allowbreak \left( 2\pi /L_{\mathrm{QPM}}\right) \mathbf{e}_{z}$, aligned
with the propagation direction, $\mathbf{e}_{z}$, to the relation between
the FF\ and SH\ wave vectors, $\mathbf{k}_{\mathrm{FF}}$ and $\mathbf{k}_{%
\mathrm{SH}}$, which may be used to cancel the original mismatch, $2\mathbf{k%
}_{\mathrm{FF}}-\mathbf{k}_{\mathrm{SH}}$. In terms of this technique, the
MM may be implemented by imposing a long-period supermodulation on the QPM
poling.

Besides its direct relevance to optics, the MM may also be implemented in
the above-mentioned atomic-molecular BEC, through the Feshbach-management
technique. As mentioned above, in terms of BEC, the application of the
latter technique, which is based on the Feshbach resonance driven by a
modulated magnetic field, to the stabilization of various types of 1D \cite%
{FRM}, 2D \cite{Fatkhulla} (and also 3D \cite{Warszawa}) matter-wave
solitons was theoretically studied in detail for the atomic BEC which obeys
the Gross-Pitaevskii equation with the cubic nonlinearity, but no similar
results were reported, thus far, for atomic-molecular condensates.

Although in the limit of large values of the mismatch a $\chi ^{(2)}$ system
may be reduced to a $\chi ^{(3)}$ limit by means of the well-known cascading
approximation, there is a fundamental difference between the systems. While $%
\chi ^{(3)}$ models may have exact Bethe-ansatz solutions \cite{LH89} even
beyond the MF approximation, the introduction of the SH (the molecular
field) lifts the integrability \cite{YBO06}. Thus, we expect effects of MM
in $\chi ^{(2)}$ systems, at small or moderate values of the mismatch, to be
different from earlier studied effects of the nonlinearity management in the
$\chi ^{(3)}$ model (in the cascading limit, the $\chi ^{(2)}$ MM goes over
into the $\chi ^{(3)}~$nonlinearity management).

The objective of the paper is to study solitons and their stability in
one-dimensional MM systems, in both the optical and BEC realizations. The MF
model, based on a set of partial differential equations, is introduced in
Sec. II. A piecewise-constant periodic modulation of the mismatch parameter
in this model is natural in the optical setting. Basic results for solitons
in the MF model are collected in Sec. III. We report stability regions for
the solitons, and conditions necessary for their self-trapping from input
beams. Interactions between the solitons are considered too, with a
conclusion that characteristics of the interactions in the MM system are
virtually the same as in its ordinary (unmodulated) counterpart.

In Sec. IV, we employ the parametric approximation \cite{YB03}, which goes
beyond the mean field, and is based on a set of ordinary differential
equations, in the case of spatially uniform configurations. Although this
model is unable to generate solitons, it allows us to analyze effects of
quantum fluctuations, which may be important in both optical \cite{HY00}
and, especially, BEC \cite{YB03} realizations of the $\chi ^{(2)}$
interactions, as well as relaxation effects, which may appear in the BEC due
to inelastic collisions. In the framework of this analysis, we adopt the
usual, in terms of the BEC, harmonic form of the periodic modulation of the
magnetic field which tunes the Feshbach resonance, rather than the
piecewise-constant format, adopted in the optical model. Numerical solutions
of parametric-approximation equations demonstrate that the non-condensate
component in the atomic-molecular gas is not essentially excited by the MM,
under a natural condition that the modulation frequency is low enough. On
the other hand, the modulation at higher frequencies may lead to destruction
of the condensate. The paper is concluded by Sec. V.

\section{The mismatch-management model for the optical medium}

In a normalized form, which is widely adopted in nonlinear optics, the
fundamental $\chi ^{(2)}$ model in one dimension is based on a system of
coupled equations for complex local amplitudes of the FF and SH\ waves, $%
u(z,x)$ and $v(z,x)$ \cite{chi2}:

\begin{equation}
\begin{array}{c}
iu_{z}+u_{xx}-u+vu^{\ast }=0, \\
2iv_{z}+v_{xx}-\alpha v+(1/2)u^{2}=0,%
\end{array}
\label{1}
\end{equation}%
where $\alpha $ is the mismatch parameter [this coefficient is an
irreducible one in the framework of the notation adopted in Eqs. (\ref{1}),
if solitons are intended to be looked for in the $z$-independent form], and
the asterisk stands for the complex conjugation. Here, it is assumed that,
in the case of spatial solitons, light propagates along axis $z$ in a planar
waveguide with the transverse coordinate $x$. In terms of temporal solitons,
$x$ is the reduced-time variable. MF equations for atomic-molecular BEC in
an atomic waveguide, where $x$ is directed along the waveguide axis, can be
reduced to the form of Eq.\ (\ref{1}) as well, with $z$ being time (see
Sec.\ \ref{nonMF} below).

In the notation of Eq.\ (\ref{1}), complete matching is attained at $\alpha
=4$, while the single exact analytical solution for the $\chi^{(2)}$ soliton
is available at $\alpha =1$,%
\begin{equation}
u=\pm \left( 3/\sqrt{2}\right) \mathrm{sech}^{2}(x/2),\,v=(3/2)\mathrm{sech}%
^{2}(x/2)\,  \label{Karamzin}
\end{equation}%
(the \textit{Karamzin-Sukhorukov} soliton \cite{SKaramzin}). At other values
of $\alpha $, solitons were found numerically, as well as in an approximate
analytical form, by means of the variational method \cite{chi2}.

The simplest model of the MM (similar, in particular, to that of the
dispersion management \cite{DM,book}) assumes periodic modulation of $\alpha
$ according to the following \textit{map} (the latter term follows the
pattern of the ``dispersion-management map" \cite{DM,book}):

\begin{equation}
\alpha (z)=\left\{
\begin{array}{ll}
\alpha _{0}-\Delta \alpha , & nL<z~<(n+1/2)L \\
\alpha _{0}+\Delta \alpha , & (n+1/2)L<z~<(n+1)L%
\end{array}%
\right. ,\,n=0,1,2,3...,  \label{2}
\end{equation}%
where $\alpha _{0}$ is the average value of $\alpha $, while $\Delta \alpha $
and $L$ are the amplitude and period of the periodic management. We here
assume equal lengths, $L/2$, of the two segments forming the MM cell.
Simulations supplementing those reported below demonstrate that a change of
the relative length of the two segments (with the respective change of the
local values of the mismatch in them) produces little effect on eventual
results, quite similar to what is known about the dispersion management \cite%
{DM,book}.

While the MM map in the piecewise-constant form of Eq. (\ref{2}) is most
natural in terms of optical waveguides, for the application to the
atomic-molecular BEC a more natural choice is, as mentioned above, the
harmonic modulation of the mismatch, corresponding to the periodic time
dependence of the magnetic field tuning the Feshbach resonance, see Eq. (\ref%
{harmonic}) below. The experience gained in the studies of various models of
the dispersion management \cite{book} suggests that the piecewise-constant
and harmonic formats of the modulation cannot lead to qualitatively
different results.

Equations (\ref{1}) conserve a known integral of motion, viz., \textit{%
Manley-Rowe invariant},
\begin{equation}
I_{\mathrm{MR}}=\int_{-\infty }^{+\infty }\left[ \left\vert u(x)\right\vert
^{2}+4\left\vert v(x)\right\vert ^{2}\right] dx,  \label{MR}
\end{equation}%
even in the case of the $z$-dependent mismatch parameter, whereas
the Hamiltonian corresponding to Eqs. (\ref{1}) is not conserved.
In addition, in the case of ``walking" solitons \cite{walking} [in
fact, these are spatial solitons tilted in the plane of $\left(
x,z\right) $],
Eqs. (\ref{1}) with variable $\alpha (z)$ conserve the total momentum,%
\begin{equation}
P=2i\int_{-\infty }^{+\infty }\left( u_{x}^{\ast }u+2v_{x}^{\ast }v\right)
dx.  \label{P}
\end{equation}%
It is also worthy to note that Eqs. (\ref{1}) with $\alpha =\alpha (z)$ are
invariant with respect to the spatial Galilean boost, hence a generic tilted
soliton, $\left( \tilde{u},\tilde{v}\right) ,$ can be generated from the
straight one by means of the corresponding transformation,%
\[
\tilde{u}(x,z)\equiv e^{i\left( c^{2}/4\right) z+i\left( c/2\right)
x}u(x-cz),~\tilde{v}(x,z)\equiv e^{i\left( c^{2}/2\right) z+icx}v(x-cz),
\]%
with an arbitrary real tilt parameter $c$. As follows from Eqs. (\ref{MR}) -
(\ref{P}), the momentum of the tilted soliton is $P=cI$, i.e., $c$ and $I$
play the role of the effective velocity and mass of the soliton.

In the case of the atomic-molecular BEC mixture, the Galilean invariance has
its literal meaning (in the temporal, rather than spatial, domain). However,
it is broken if the axial trapping potential is taken into regard.

\section{Dynamics of spatial solitons under the mismatch management in the
optical model}

\subsection{Formation of stable mismatch-managed solitons}

To simulate the transmission of spatial optical solitons under the MM
conditions, we solved equations (\ref{1}) with MM map (\ref{2}) by means of
the split-step numerical method, which uses the Fourier transform to handle
the linear stage of the evolution. As the input (initial pulse), we took
either an ordinary soliton corresponding to the averaged version of the
model, i.e., one with $\alpha (z)\equiv \alpha _{0}$, or a deliberately
altered pulse, to verify whether the MM system will provide for
self-trapping into a stable transmission regime. Below, we display
systematic results obtained for $\alpha _{0}=1$ and $\alpha _{0}=2$
(comparison with results collected with other values of the average mismatch
parameter demonstrate that these two cases adequately represent the generic
situation). In particular, for $\alpha _{0}=1$ we launched the initial pulse
taken as exact solution (\ref{Karamzin}) corresponding to $\alpha =1$. For $%
\alpha _{0}=2$, we typically used either a soliton solution found in a
numerical form for $\alpha =2$, or, in order to try the effect of a strong
change of the input, we again took expression (\ref{Karamzin}), i.e., the
exact soliton appertaining to $\alpha =1$. At a fixed value of $\alpha _{0}$%
, results were collected by varying the MM amplitude $\Delta \alpha $ and
period $L$.

\begin{figure}[tbp]
\centering\subfigure[]{\includegraphics[width=3.0in]{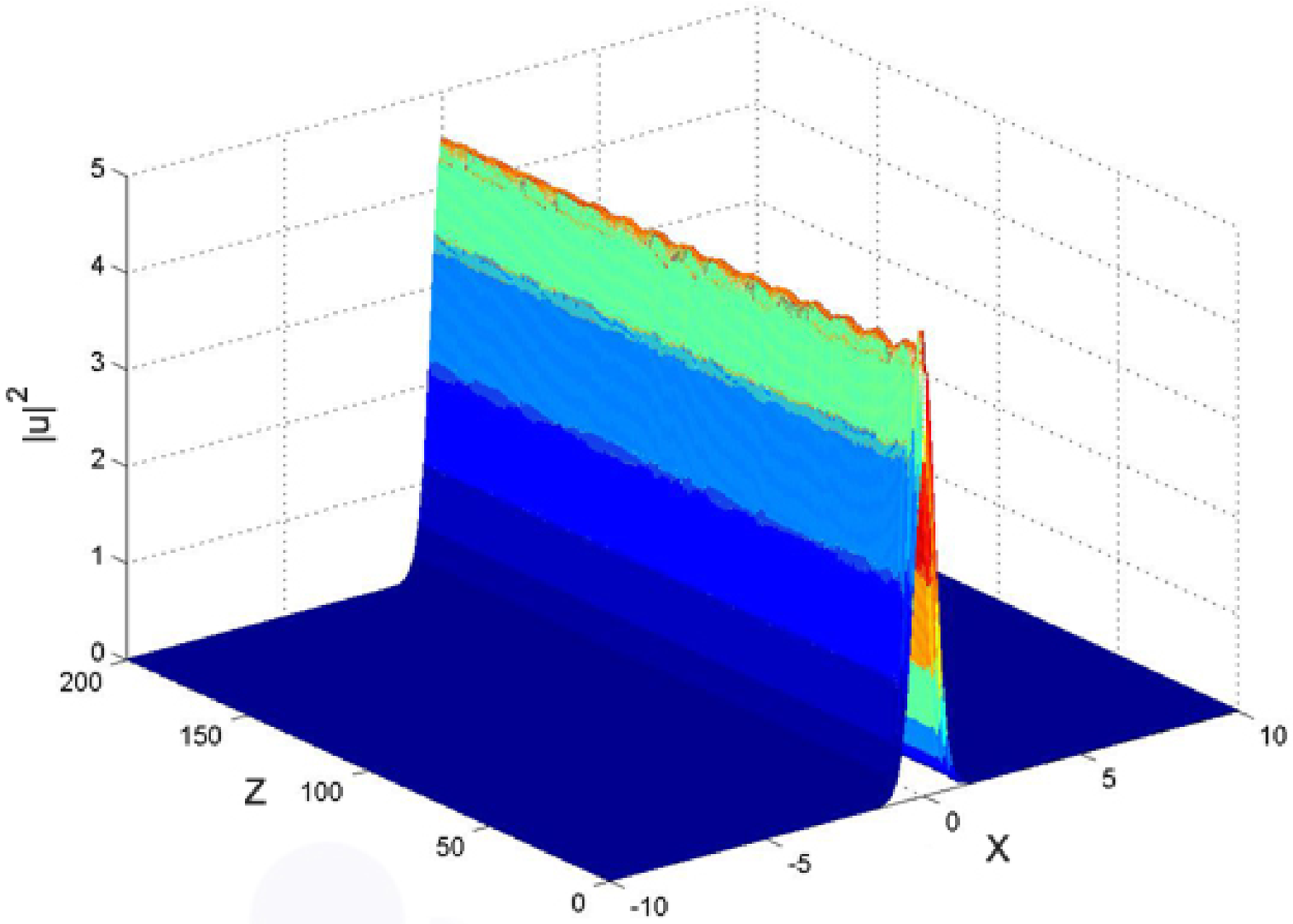}} \centering%
\subfigure[]{\includegraphics[width=3.0in]{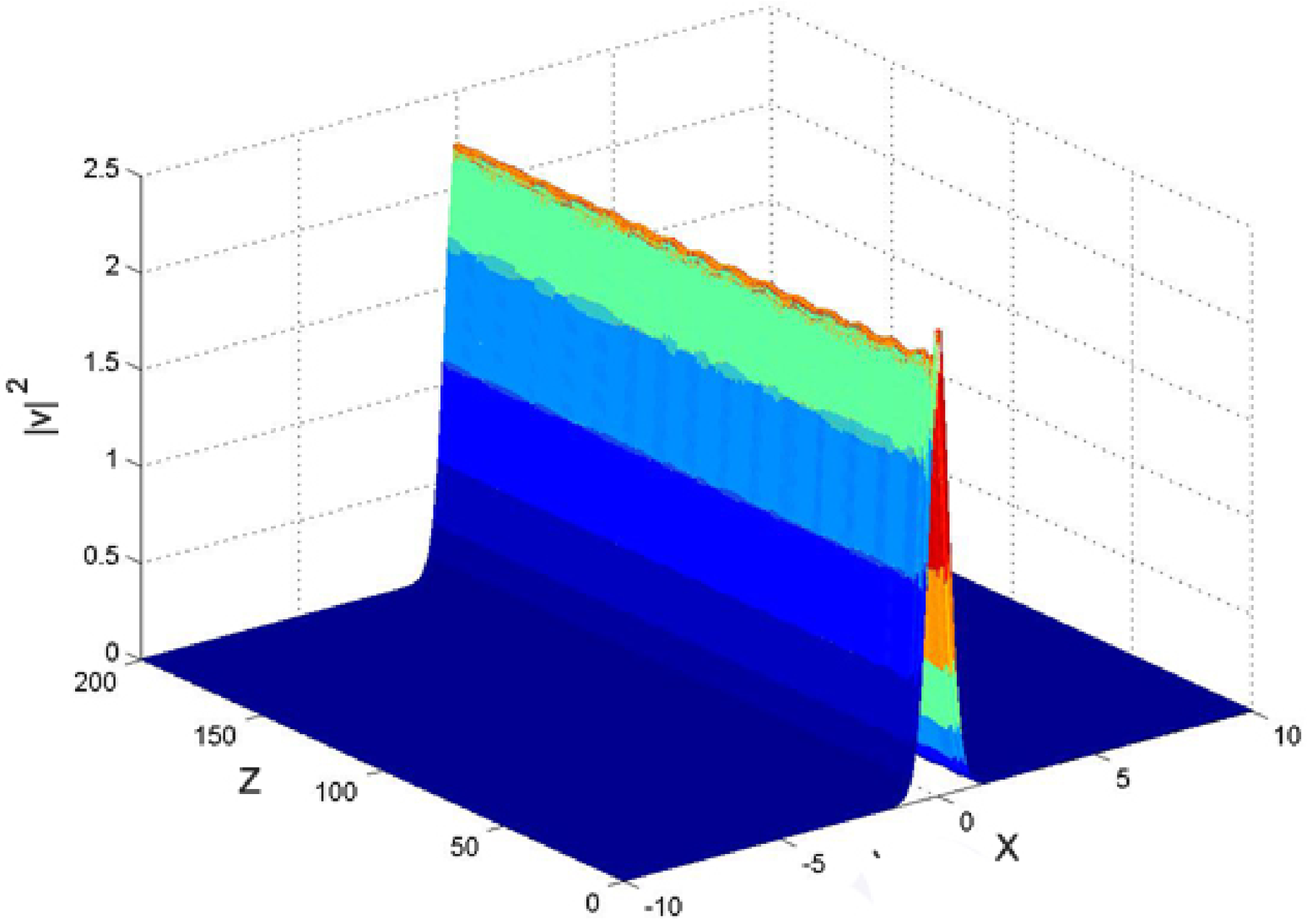}}
\caption{(Color online) A typical example of the quick
self-trapping of an initial beam into a stable spatial soliton in
the mismatch-management model, with $\protect\alpha _{0}=L=\Delta
\protect\alpha =1$. The input was the
\textit{Karamzin-Sukhorukov} soliton \protect\cite{SKaramzin} for $\protect%
\alpha (z)\equiv 1$, taken as per Eq. (\protect\ref{Karamzin}).$_{.}$Panels
(a) and (b) show the evolution of the FF and SH fields.}
\label{Fig1}
\end{figure}

First, in Fig. \ref{Fig1} we display a typical numerical solution for $%
\alpha _{0}=1$, $L=1$ and $\Delta \alpha =1$, generated with the use of
exact soliton (\ref{Karamzin}) as the initial condition. In this figure,
panels (a) and (b) present the evolution of the FF and SH components of the
field. In fact, in all cases considered, there was no conspicuous difference
in the dynamics of the two components, therefore in other cases shown below
we only display the picture for the FF beam. As is seen from Fig. \ref{Fig1}%
, the input beam readily gives rise to a robust spatial soliton (intrinsic
pulsations with period $L=1$, caused by the MM, are almost invisible in Fig. %
\ref{Fig1}); the soliton remains stable in indefinitely long simulations.
The transient stage, necessary for the self-trapping, is fairly short,
comprising a few MM cells. In the subsequent evolution, gradually fading
residual oscillations of the pulse's amplitude can be seen, with a period
covering several cells (these oscillations are caused by the initial
perturbation, rather than the periodic MM). Further examples of stable
transmission regimes are displayed below in Figs. \ref{Fig3}(a) and (c).

\subsection{Stability diagrams}

\begin{figure}[tbp]
\centering\subfigure[]{\includegraphics[width=3.0in]{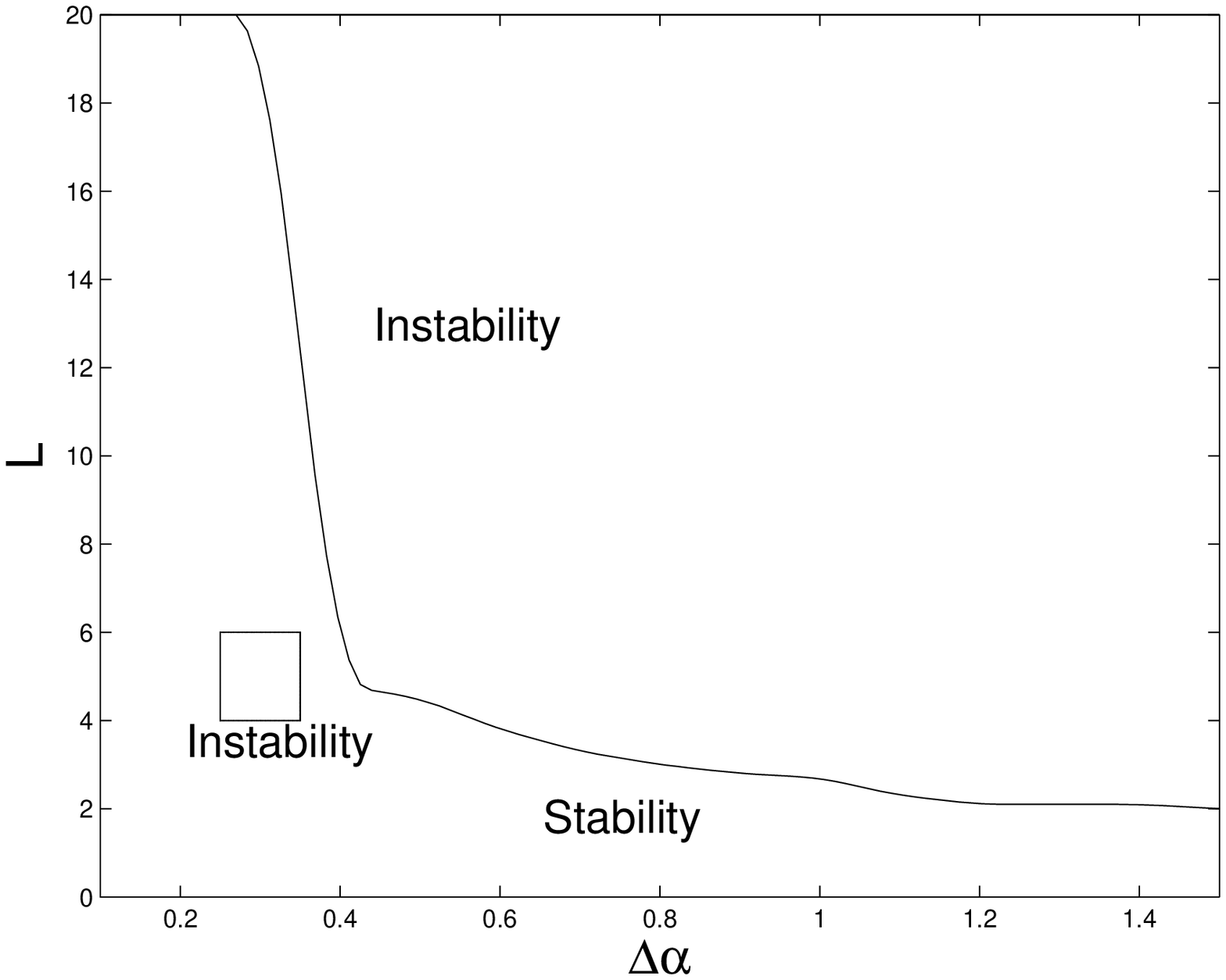}} \centering%
\subfigure[]{\includegraphics[width=3.0in]{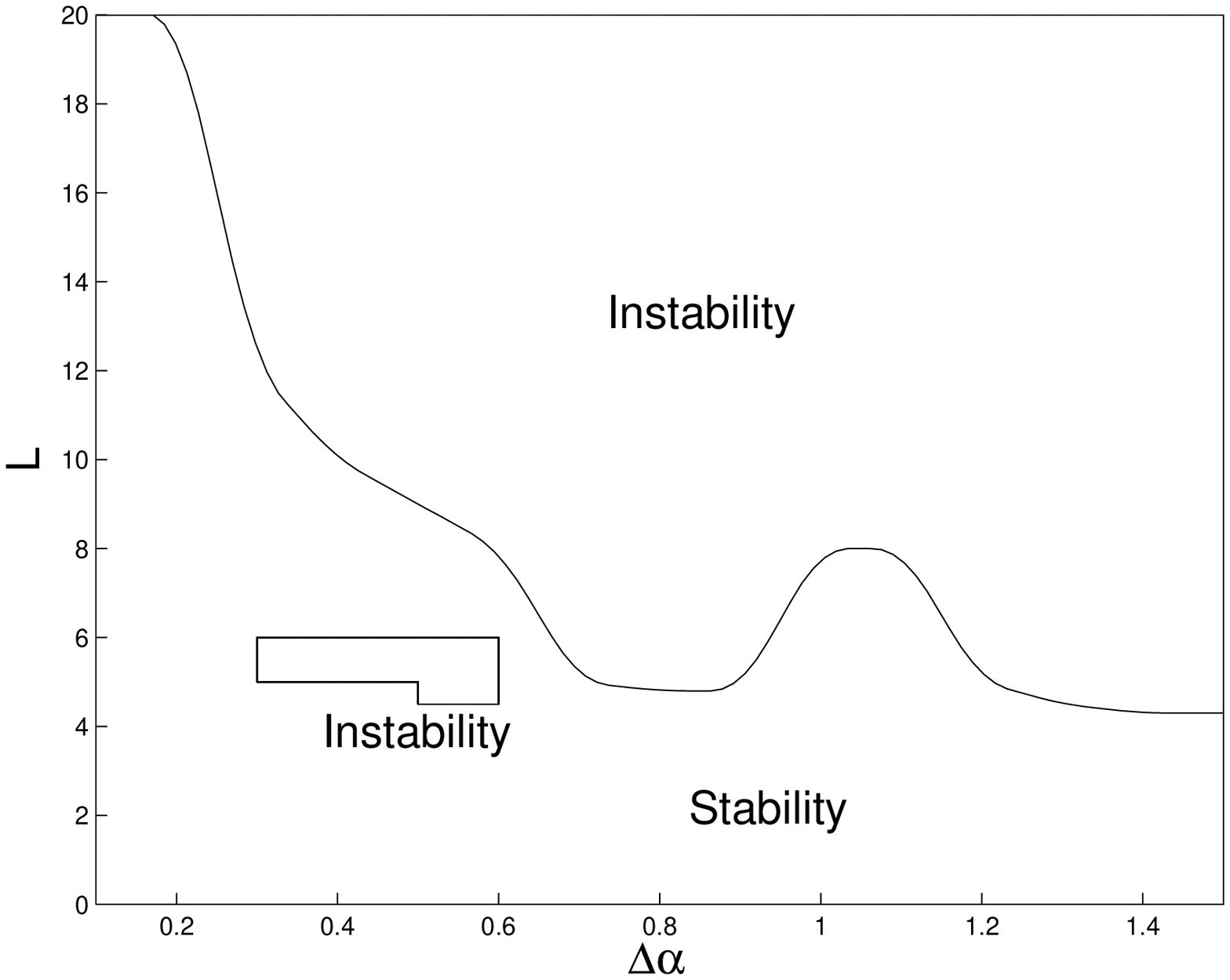}}
\caption{Stability and instability areas in the
mismatch-management model, in the plane of the management
amplitude ($\Delta \protect\alpha $) and period ($L$). Panels (a)
and (b) display the stability diagrams for the
model with average mismatch $\protect\alpha _{0}=1$ and $\protect\alpha %
_{0}=2$, respectively. In either case, the diagram was built by collecting
results of many simulations, with the input taken as a soliton of the
respective averaged model, i.e., one with $\protect\alpha (z)\equiv \protect%
\alpha _{0}$ [for $\protect\alpha _{0}=1$, the initial pulse is the
analytical Karamzin-Sukhorukov solution, given by Eq. (\protect\ref{Karamzin}%
)].}
\label{Fig2}
\end{figure}

Conclusions drawn from systematic simulations are summarized in the
stability diagrams, which are displayed in Fig. \ref{Fig2} for $\alpha
_{0}=1 $ (a) and $\alpha _{0}=2$ (b). As said above, in the former case we
launched the pulse corresponding to Karamzin-Sukhorukov soliton (\ref%
{Karamzin}), which is an exact solution for $\alpha (z)\equiv 1$, and in the
latter case the initial pulse was a numerically found stationary soliton
corresponding to $\alpha (z)\equiv $ $2$. The diagrams display areas in
parameter plane $\left( \Delta \alpha ,L\right) $ where the initial pulse
gives rise to stable transmission, or decay of the pulse [an example of the
latter outcome is displayed below in Fig. \ref{Fig3}(b)]. Naturally, the
stability regions tend to extend along the parameter axes, as, in either
limit of $\Delta \alpha \rightarrow 0$ or $L\rightarrow 0$, the model
returns to the usual $\chi ^{(2)}$ system (in the case of $L\rightarrow 0$,
this is provided by averaging), where the initial pulse represents an
ordinary stable soliton.

Generally, the instability of the $\chi ^{(2)}$ solitons in a part of the
parameter space may be realized as a result of a \textit{resonance} between
the perturbation frequency, introduced by the periodic action of the MM, and
the frequency of the intrinsic mode, which, as is well known, $\chi ^{(2)}$
solitons have in the system with constant coefficients [on the contrary to
the nonlinear-Schr\"{o}dinger (NLS) solitons] \cite{chi2}. Indeed,
comparison of the numerically found instability border, shown in Figs. \ref%
{Fig2}(a) and (b), with values of the intrinsic eigenfrequency, which are
known from numerical computations too, demonstrates that the explanation of
the instability by the resonance is feasible, although the intrinsic
frequency is defined for infinitely small perturbations, while the
instability sets in when the solitons are strongly perturbed.

It is also relevant to note that, in the limit of large positive values of
mismatch $\alpha $, the SH (or the molecular MF, in terms of BEC) can be
eliminated from Eqs.\ (\ref{1}) by means of the well-known cascading
approximation, $v\approx u^{2}/2\alpha $, reducing the $\chi ^{(2)}$ system
to the single NLS equation. The same is valid when the large mismatch is
subjected to the management [provided that $\alpha (z)$ does not change its
sign]. In that case, the cascading limit will lead (as mentioned above) to
the NLS equation featuring periodic \textit{nonlinearity management} \cite%
{book}. In fact, solitons in the latter equation and their stability were
investigated in some detail in various contexts \cite{FRM,book}. In
particular, a resonant mechanism of the destabilization of higher-order
solitons (bound states of fundamental solitons) under the action of the
nonlinearity management, qualitatively similar to one outlined above, has
been demonstrated in Ref. \cite{HS}.

A noteworthy feature of the diagrams is the presence of an \textit{%
instability enclave} inside the stability area. In Fig. \ref{Fig2}(a), the
enclave is shown symbolically by a square, as exact delineation of its
borders requires extremely long simulations. In Fig. \ref{Fig2}(b), the
borders of the enclave approximately correspond to its real shape. It may be
relevant to mention that examination of similar stability diagrams in the
above-mentioned SSM had revealed a different but somewhat similar feature,
viz., a system of \textit{stability islands} inside the instability area at
large values of $L$ \cite{SSM}. In the present MM model, the simulations do
not reveal stability islands (on the other hand, no ``instability lakes",
that would be similar to the enclaves in Fig. \ref{Karamzin}, were found
inside the stability area in the SSM). We surmise that there may exist
additional small instability enclaves, and, in principle, they may even form
a fractal pattern. However, an accurate investigation of these issues
requires extremely high numerical accuracy, and they are left beyond the
scope of this work.

The existence of the instability enclave is further illustrated by a set of
simulations presented in Fig. \ref{Fig3}, which are performed along a
vertical line, $\Delta \alpha =0.3$, cutting through the enclave in Fig. \ref%
{Fig2}(a). A notable difference is observed in the self-trapping into stable
transmission regimes below and above the instability region: in the former
case, at $L=4$ (which is close to the instability border), the established
pulse is very different from the input, due to considerable radiation loss
in the process of the establishment of the stable-transmission regime (in
the SH component of the wave field, which is not shown in Fig. \ref{Fig3},
an approximately the same degree of the loss is observed). It is relevant to
mention that the formation of stable solitons in the SSM may also be
accompanied by strong losses, depending on parameters of the system and the
form of the input \cite{SSM}. On the other hand, above the instability
enclave, the loss is small, and the established soliton is closer to the
input, as seen in Fig. \ref{Fig3}(c) (the same is observed in the SH
counterpart of the latter figure, which is not shown here).
\begin{figure}[tbp]
\centering\subfigure[]{\includegraphics[width=3.0in]{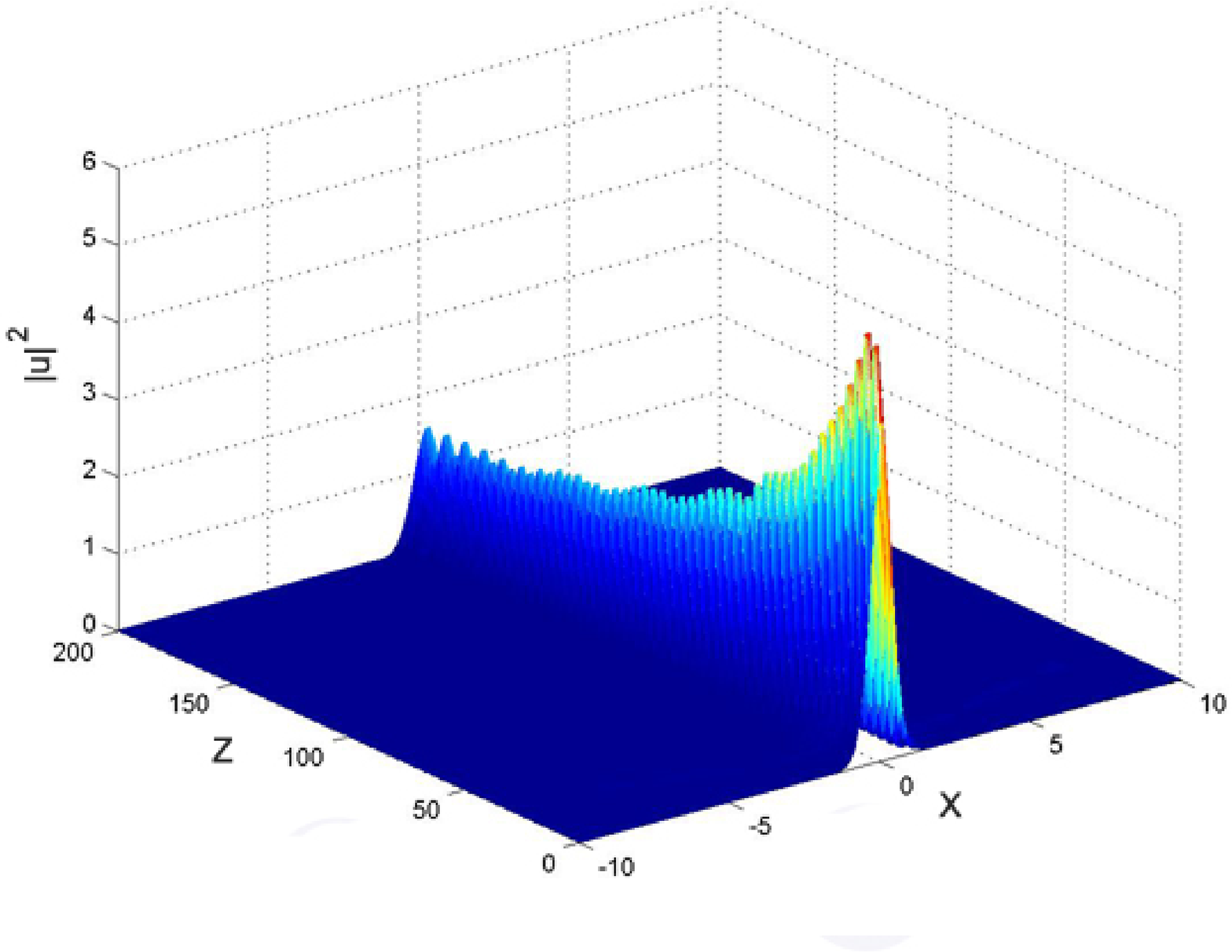}} %
\subfigure[]{\includegraphics[width=3.0in]{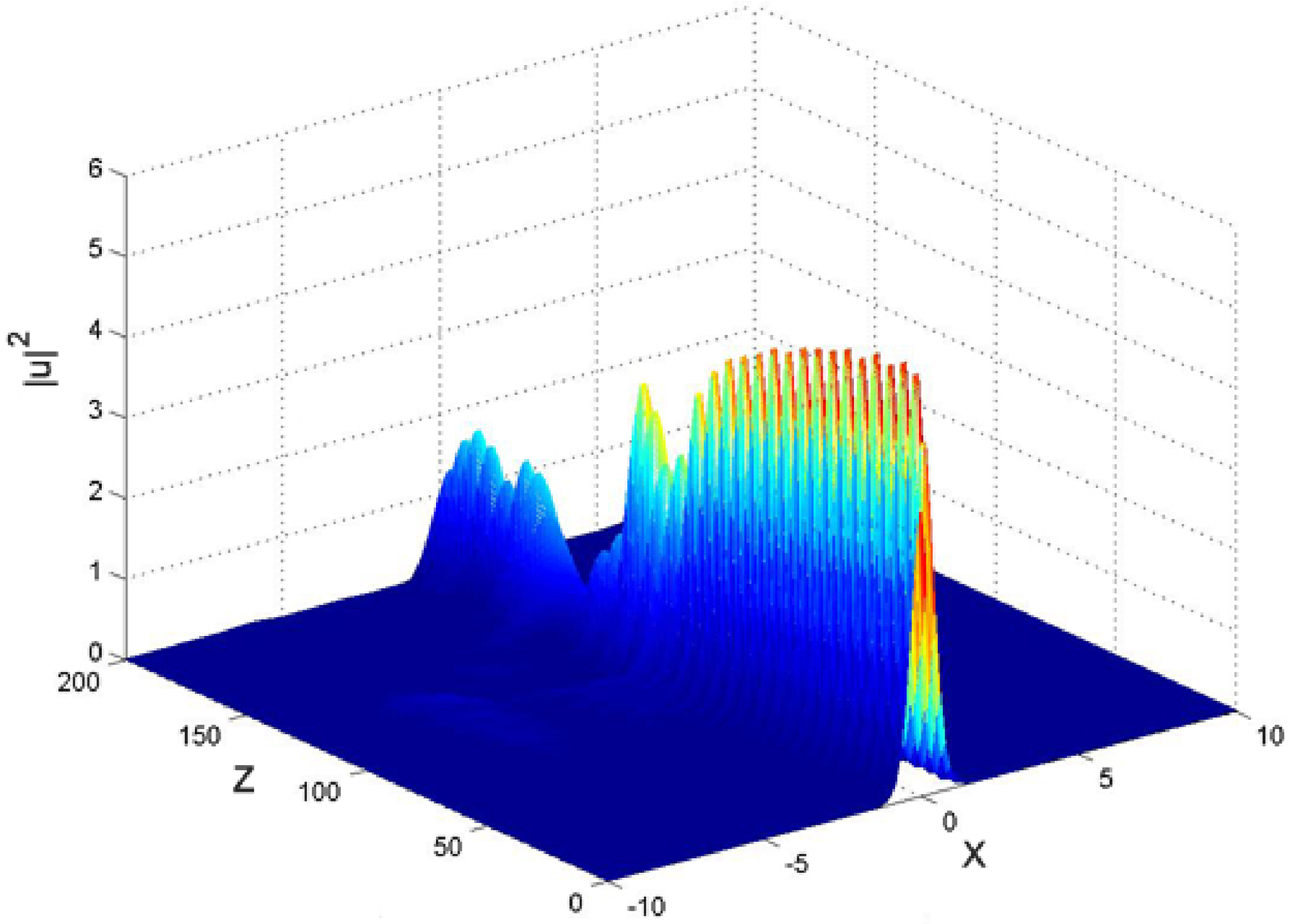}}\centering %
\subfigure[]{\includegraphics[width=3.0in]{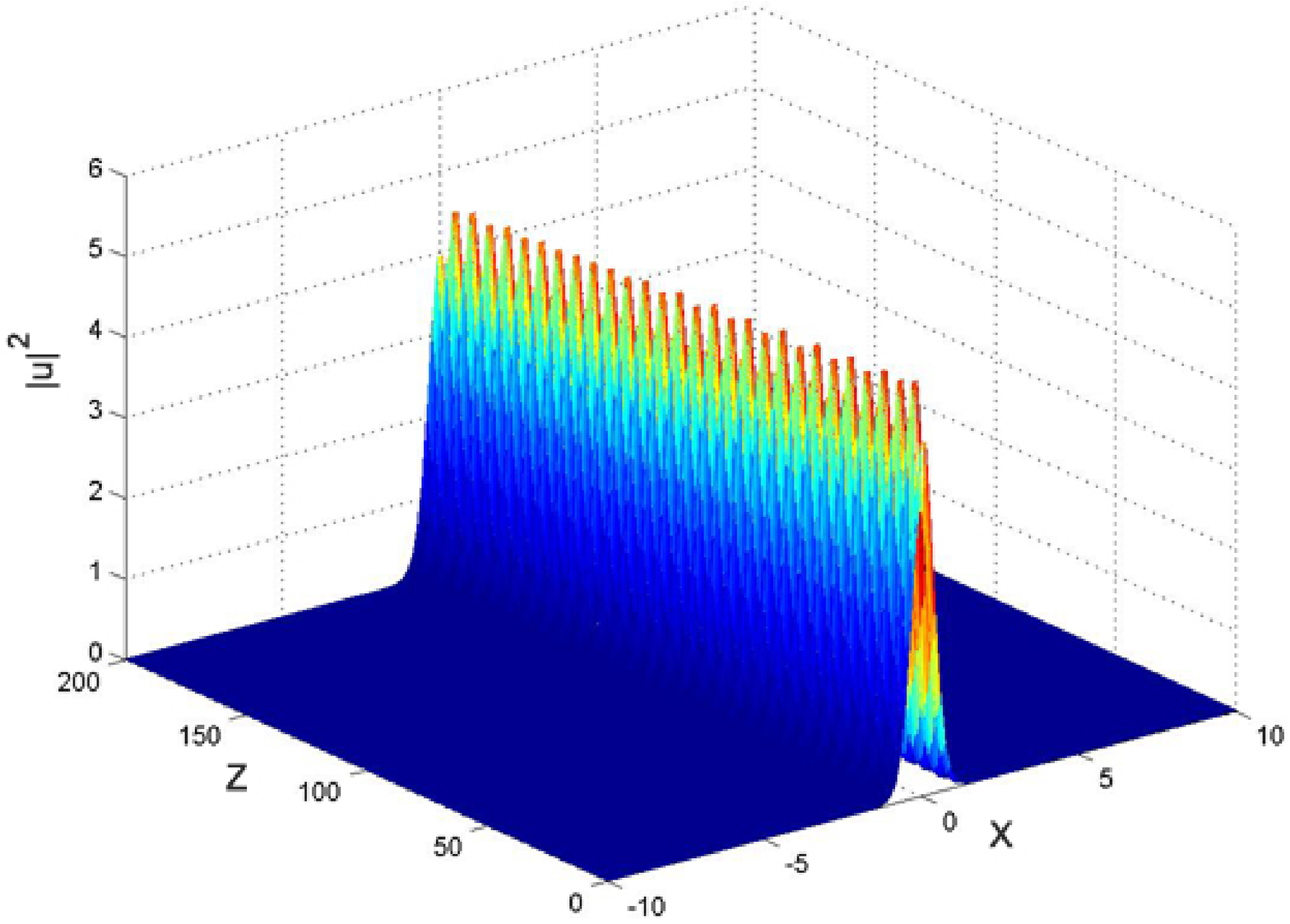}}\centering
\caption{(Color online) Three examples of the evolution generated
by the input in the form of Eq. (\protect\ref{Karamzin}) launched
into the
mismatch-management system with $\protect\alpha _{0}=1$ and $\Delta \protect%
\alpha =0.3$, the modulation period being $L=4$ (a), $6$ (b), and $7$ (c).
Point (b) falls into the instability enclave in Fig. \protect\ref{Fig2}(a);
in this case, the soliton suffers, eventually, complete destruction. }
\label{Fig3}
\end{figure}

\subsection{The system's tolerance to variations of the input pulse}

Another noteworthy feature revealed by the simulations is great tolerance of
the spatial solitons established in the MM system to variance of the input
beam. In particular, if the input launched into the system with $\alpha
_{0}=2$ was deliberately taken in the ``wrong" form of expressions (\ref%
{Karamzin}) (recall they would yield an exact solution for the averaged
equations with $\alpha =1$, rather than $\alpha =2$), the simulations,
performed for various values of $\Delta \alpha $ and $L$, produced a
stability diagram nearly identical to that shown in Fig. \ref{Fig2}(b).

\begin{figure}[tbp]
\centering\subfigure[]{\includegraphics[width=3.0in]{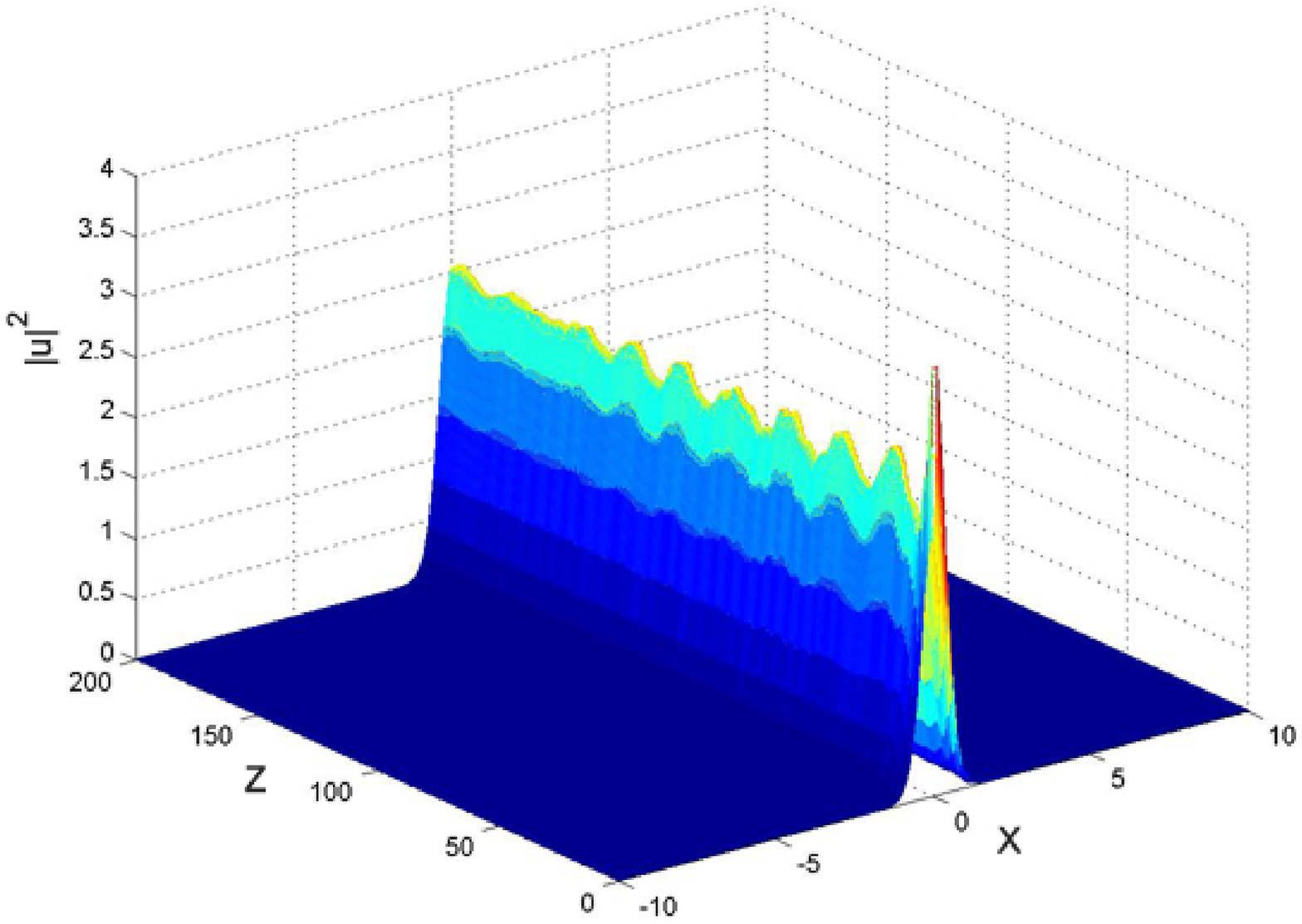}} \centering%
\subfigure[]{\includegraphics[width=3.0in]{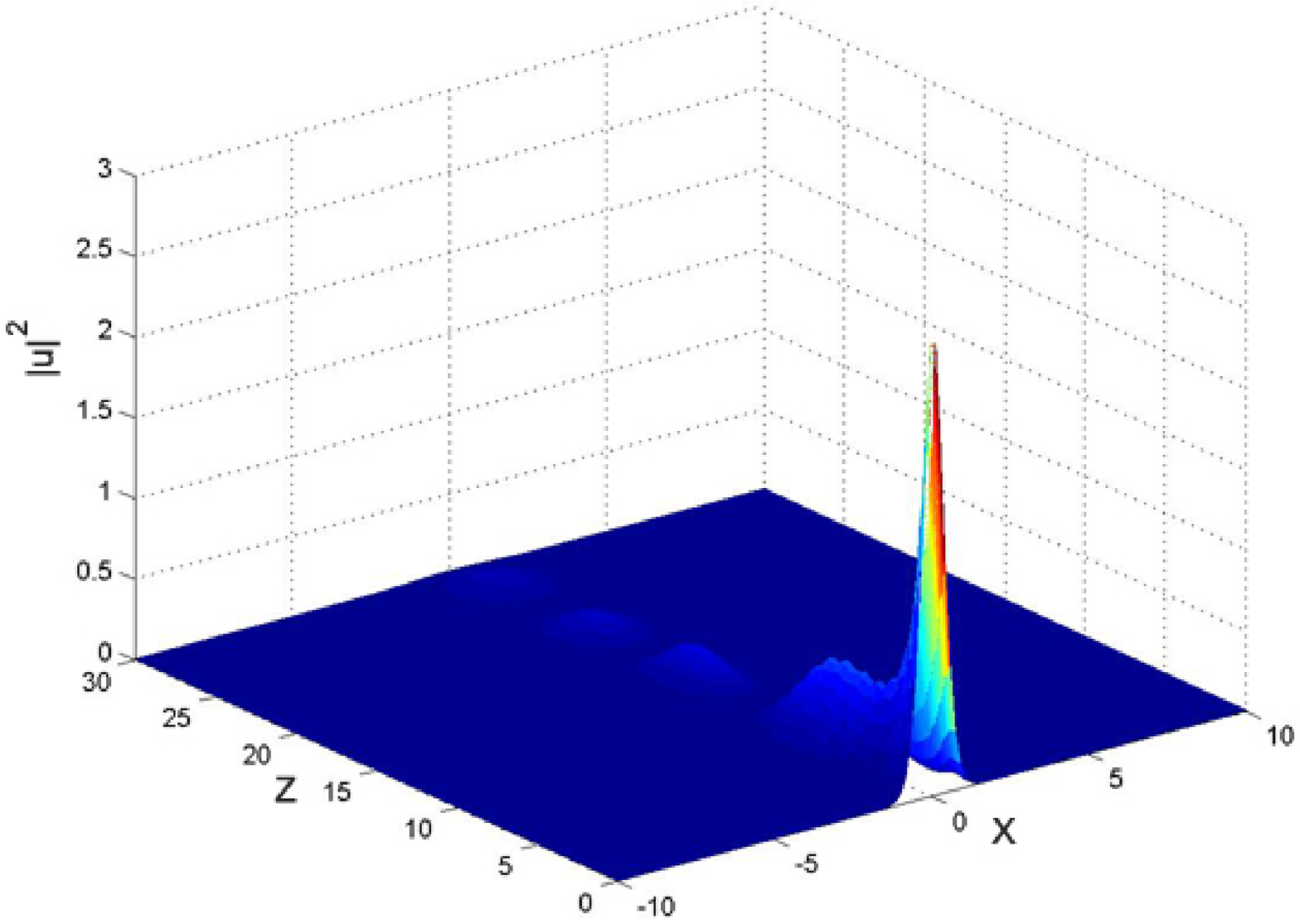}}
\caption{(Color online) (a) Self-trapping into a stable
mismatch-managed
spatial soliton of the input taken as per Eq. (\protect\ref{W}) with $W=0.81$%
. (b) Decay of the input beam taken as in Eq. (\protect\ref{W}), but with $%
W=0.64$. In either case, $u_{0}(x)$ and $v_{0}(x)\ $are components of
soliton solution (\protect\ref{Karamzin}), and parameters are $\protect%
\alpha _{0}=\Delta \protect\alpha =$ $L=1$.}
\label{Fig4}
\end{figure}

It is natural to expect that decrease of the input power will eventually
lead to a failure in the self-trapping of the spatial soliton, i.e., there
must exist a certain power threshold for the soliton formation. To find it,
we performed additional simulations with the inputs similar to those used
above [in particular, waveform (\ref{Karamzin}) was taken for $\alpha _{0}=1$%
], but multiplied by a power-reducing factor, $W<1$:
\begin{equation}
\left\{ u_{0}(x),v_{0}(x)\right\} \rightarrow \sqrt{W}\left\{
u_{0}(x),v_{0}(x)\right\} .  \label{W}
\end{equation}%
It was found that the critical value of $W$, below which the thus altered
input pulse (\ref{Karamzin}) fails to generate a stable soliton is $W_{%
\mathrm{cr}}\approx 0.72$ (in fact, it is close to a critical value that can
be found numerically in the ordinary $\chi ^{(2)}$ system, with $\Delta
\alpha =0$). Figures \ref{Fig4}(a,b) show what happens when $W$ is taken,
respectively, above and below $W_{\mathrm{cr}}$.

It was also checked that the change of the distribution of the Manley-Rowe
invariant [see Eq. (\ref{MR})] between the two components of the input beam
virtually does not affect the self-trapping threshold (in particular, one
may start with the entire power put in the FF field, while $v_{0}=0$). The
ordinary $\chi ^{(2)}$ system, with $\Delta \alpha =0$, features a similar
property \cite{chi2}.

\subsection{Soliton-soliton interactions}

\begin{figure}[tbp]
\centering{\includegraphics[width=5in]{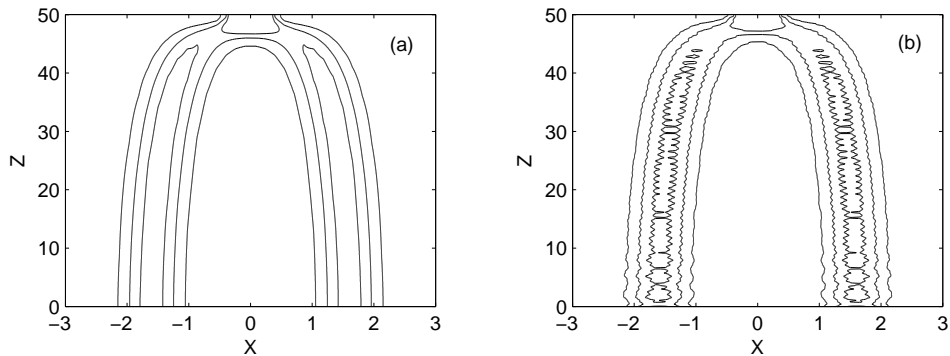}}
\caption{Interaction of two identical in-phase solitons in the ordinary $%
\protect\chi ^{(2)}$ model (with $\protect\alpha =1,~\Delta \protect\alpha %
=0 $) (a), and its mismatch-managed counterpart with $\protect\alpha %
_{0}=\Delta \protect\alpha =L=1$ (b). The initial separation between the
solitons is $\Delta x=3.8$, while the minimum separation at which the
solitons do not interact is $\Delta x_{\min }\approx 5$, in this case.}
\label{Fig5}
\end{figure}

We have also performed systematic simulations of interactions between
solitons in the MM model. It was found that the character of the interaction
remains virtually the same as in the ordinary $\chi ^{(2)}$ model with
constant mismatch, that was studied in detail in earlier works \cite{chi2}.
The main characteristic of the interaction is a minimum initial separation
between co-propagating identical in-phase solitons, $\Delta x_{\min }$,
which is defined so that the solitons do not demonstrate any interaction for
$\Delta x>\Delta x_{\min }$, while, being placed at distance $\Delta
x<\Delta x_{\min }$, they start to attract each other and eventually merge
into a single beam. A typical example displayed in Fig. \ref{Fig5} shows
that the interaction indeed seems identical in the MM system and its
ordinary counterpart.

\section{Effects of relaxation and quantum fluctuations}

\label{nonMF}

As explained in Introduction, the $\chi ^{(2)}$ model can also describe a
hybrid atom-molecule BEC \cite{TTHK99} under quasi-1D confinement. In the
parametric approximation, developed for the case of free space in Ref.\ \cite%
{YB03}, the condensate is described by the following system of equations of
motion for the molecular mean field $\varphi _{m}\left( X,t\right) $ and the
atomic annihilation operators $\hat{\Psi}_{a}\left( X,t\right) $:
\begin{eqnarray}
i{\frac{\partial }{dt}}\varphi _{m}\left( X,t\right) &=&-{\frac{1}{4m}}{%
\frac{\partial ^{2}}{\partial X^{2}}}\varphi _{m}\left( X,t\right) +g\langle
\hat{\Psi}_{a}\left( X,t\right) \hat{\Psi}_{a}\left( X,t\right) \rangle
\nonumber \\
&&-i\left( {\frac{k_{a}}{2}}\langle \hat{\Psi}_{a}^{\dag }\left( X,t\right)
\hat{\Psi}_{a}\left( X,t\right) \rangle +k_{m}|\varphi _{m}\left( X,t\right)
|^{2}\right) \varphi _{m}\left( X,t\right) ,  \label{phim} \\
i{\frac{\partial }{dt}}\hat{\Psi}_{a}\left( X,t\right) &=&\left[ -{\frac{1}{%
2m}}{\frac{\partial ^{2}}{\partial X^{2}}}-{\frac{1}{2}}D_{1D}\left(
t\right) \right] \hat{\Psi}_{a}\left( X,t\right)  \nonumber \\
&&+2g^{\ast }\varphi _{m}\left( X,t\right) \hat{\Psi}_{a}^{\dag }\left(
X,t\right) -\frac{1}{2}ik_{a}|\varphi _{m}\left( X,t\right) |^{2}\hat{\Psi}%
_{a}\left( X,t\right) +i\hat{F}\left( X,t\right) ,  \label{Psia}
\end{eqnarray}%
where $m$ is the atomic mass and units with $\hbar =1$ are used. As shown in
Ref. \cite{Y05}, the 1D coupling constant, $g$, and (time-dependent)
detuning $D_{1D}$ can be expressed, as follows, in terms of elastic
scattering length $a_{\mathrm{bg}}$, phenomenological Feshbach-resonance
strength $\Delta $, the difference between the magnetic momenta of an atomic
pair in the open and closed channels, $\mu $, detuning of the external
variable magnetic field, $B(t)$, from its resonant value $B_{0}$, and
transverse trap frequency $\omega _{\perp }$:
\begin{equation}
|g|^{2}=\omega _{\perp }|a_{\mathrm{bg}}\mu |\Delta ,\qquad D_{1D}\left(
t\right) =\mu \left[ B\left( t\right) -B_{0}\right] -\omega _{\perp }~.
\label{gD1D}
\end{equation}%
The effect of the confinement-induced resonance \cite{O98} is neglected in
Eq.\ (\ref{gD1D}), as $a_{\mathrm{bg}}$ is much smaller than the transverse
size of the trap. The source of the quantum noise in Eq. (\ref{Psia}), $\hat{%
F}\left( X,t\right) $, and rate coefficient $k_{a}$ account for the
deactivation in atom-molecule inelastic collisions \cite{YB03}. Rate
coefficients $k_{a}$ and $k_{m}$ (for molecule-molecule collisions) are
related to their 3D counterparts, $k_{a,m}=\left( m\omega _{\perp }/2\pi
\right) k_{a,m}^{\left( 3D\right) }$.

Neglecting the collision-induced deactivating, and replacing the
atomic-field operator $\hat{\Psi}_{a}\left( X,t\right) $ by $c$-number mean
field $\varphi _{0}\left( X,t\right) $, the substitution
\begin{eqnarray}
\varphi _{0}\left( X,t\right) &=&\Phi u\exp \left( -i\int \Omega \left(
t\right) dt\right) ,\qquad \varphi _{m}\left( X,t\right) =-\sqrt{2}\Phi
v\exp \left( -2i\int \Omega \left( t\right) dt\right) ,  \nonumber \\
&&  \label{scaling} \\
t &=&z/\left( 2\sqrt{2}g\Phi\right) ,\qquad X=x/\left( 2^{5/4}\sqrt{mg\Phi}%
\right) ,\qquad D_{1D}=\sqrt{2}g\Phi\left( \alpha -4\right)  \nonumber
\end{eqnarray}%
with $\Omega \left( t\right) \equiv -D_{1D}\left( t\right) /2-2\sqrt{2}g\Phi$%
,  casts Eqs.\ (\ref{phim}) and (\ref{Psia}) precisely in the form of Eqs.\ (%
\ref{1}), which are adopted in nonlinear optics. Necessary normalization of $%
u$ and $v$ can be provided by the proper choice of the mean-field scaling
constant, $\Phi$, in Eqs. (\ref{scaling}).

Effects of deactivation and quantum fluctuations neglected in normalized
equations\ (\ref{1}) can be taken into regard in the framework of the
parametric approximation \cite{YB03}, where the $X$-dependence of $\varphi
_{m}$ is neglected, and the atomic field operator is represented as
\begin{eqnarray}
\hat{\Psi}_{a}\left( X,t\right) &=&\left( 1/\sqrt{2\pi }\right) \int
dpe^{ipx}C\left( t\right) \left[ \hat{A}\left( p,t\right) \psi _{c}\left(
p,t\right) +\hat{A}^{\dag }\left( -p,t\right) \psi _{s}\left( p,t\right) %
\right] ,  \label{Psirep} \\
C\left( t\right) &=&\exp \left( -\frac{1}{2}\int\limits_{0}^{t}dt^{\prime
}k_{a}|\varphi _{m}\left( t^{\prime }\right) |^{2}\right) .  \nonumber
\end{eqnarray}%
Here, $c$-number functions $\psi _{c,s}\left( p,t\right) $ satisfy the
time-evolution equations,
\begin{equation}
i\dot{\psi}_{c,s}\left( p,t\right) =\left[ {\frac{p^{2}}{2m}}-{\frac{1}{2}}%
D_{1D}\left( t\right) \right] \psi _{c,s}\left( p,t\right) +2g^{\ast
}\varphi _{m}\left( t\right) \psi _{s,c}^{\ast }\left( p,t\right) .
\label{Psics}
\end{equation}%
In the present analysis, the initial moment, $t=0$, corresponds to a
relatively small detuning. As the initial state is implied to be a stable
condensate, Eq.\ (\ref{Psirep}) may be considered, at $t=0$, as the
Bogoliubov transformation, with operators $\hat{A}\left( p,0\right) $ being
annihilation operators of the Bogoliubov quasiparticles. In this case,
initial conditions for functions $\psi _{c,s}(t)$ are produced by the
Bogoliubov transform,
\[
\psi _{c}\left( p,0\right) =\left( {\frac{d_{p}+\epsilon _{p}}{2\epsilon _{p}%
}}\right) ^{1/2},~\psi _{s}\left( p,0\right) =-2{\frac{g^{\ast }\varphi
_{m}\left( 0\right) }{d_{p}+\epsilon _{p}}}\psi _{c}^{\ast }\left(
p,0\right)
\]%
(for $p\neq 0$),where $d_{p}\equiv p^{2}/\left( 2m\right) -D_{1D}\left(
0\right) /2-E$, the Bogoliubov excitation energy is $\epsilon _{p}=\sqrt{%
d_{p}^{2}-4|g\varphi _{m}\left( 0\right) |^{2}}$, and $E$ is the chemical
potential of the atom-molecule condensate. At zero temperature, one has $%
\hat{A}\left( p,0\right) |$in$\rangle =\left( 2\pi \right) ^{1/2}\varphi
_{0}\left( 0\right) \delta \left( p\right) |$in$\rangle $, where $|$in$%
\rangle $ is the initial-state vector, and the atomic-condensate mean field $%
\varphi _{0}\left( t\right) $ is expressed in terms of solutions to Eq.\ (%
\ref{Psics}), which satisfy initial conditions $\psi _{c}\left( 0,0\right)
=1 $, $\psi _{s}\left( 0,0\right) =0$, as follows:
\[
\varphi _{0}\left( t\right) =\langle \text{in}|\hat{\Psi}_{a}\left(
X,t\right) |\text{in}\rangle =C\left( t\right) \left[ \psi _{c}\left(
0,t\right) \varphi _{0}\left( 0\right) +\psi _{s}\left( 0,t\right) \varphi
_{0}^{\ast }\left( 0\right) \right] .
\]%
Then, the normal and anomalous densities in Eq.\ (\ref{phim}) become $X$%
-independent, as shown in Ref.\ \cite{YB03}:
\begin{eqnarray*}
\langle \hat{\Psi}_{a}\left( X,t\right) \hat{\Psi}_{a}\left( X,t\right)
\rangle &=&|\varphi _{0}\left( t\right) |^{2}+{\frac{1}{2\pi }}%
\int\limits_{-\infty }^{+\infty }dpn_{s}\left( p,t\right) \\
\langle \hat{\Psi}_{a}^{\dag }\left( X,t\right) \hat{\Psi}_{a}\left(
X,t\right) \rangle &=&\varphi _{0}^{2}\left( t\right) +{\frac{1}{2\pi }}%
\int\limits_{-\infty }^{+\infty }dpn_{s}\left( p,t\right) ,
\end{eqnarray*}%
\noindent where the momentum distributions of non-condensate atoms, $%
n_{s}\left( p,t\right) $ and $m_{s}\left( p,t\right) $, can be expressed in
terms of $\psi _{c,s}\left( p,t\right) $ \cite{YB03}. Accordingly, the
equation for the molecular mean field, Eq. (\ref{phim}), takes the form of
\begin{eqnarray}
i\dot{\varphi}_{m}\left( t\right) &=&g\varphi _{0}^{2}\left( t\right) +{%
\frac{g}{\pi }}\int\limits_{p_{\min }}^{\infty }dpm_{s}\left( p,t\right)
\nonumber \\
&&-i\left( {\frac{k_{a}}{2}}|\varphi _{0}\left( t\right) |^{2}+{\frac{k{\ }%
_{a}}{2\pi }}\int\limits_{0}^{\infty }dpn_{s}\left( p,t\right)
+k_{m}|\varphi _{m}\left( t\right) |^{2}\right) \varphi _{m}\left( t\right) .
\label{phimf}
\end{eqnarray}

Unlike the 3D case \cite{YB03}, the 1D problem does not require
renormalization, as the integral of $m_{s}$ is free of the ultraviolet
divergence. Nevertheless, it now diverges at zero momentum. This formal
infrared divergence is related to phase fluctuations and to the absence of
true condensate in the infinite 1D system. However, in the present work we
actually consider coordinate-dependent solitons, while the divergence is a
consequence of the neglect of the coordinate dependence in the parametric
approximation. A more careful analysis of the inhomogeneous case yields an
asymptotic estimate, $m_{s}\left( p,t\right) \sim p^{2}$ for $p\ll p_{\min }$%
, where the characteristic momentum is inversely proportional to the
soliton's size, $p_{\min }=1/\sqrt{mg\Phi}$. Thus, $p_{\min }$ may be
naturally chosen as the lower-integration limit in Eq.\ (\ref{phimf}). The
resultant value of the integral features a weak dependence on the lower
limit, as the divergence is logarithmic.

The characteristic kinetic energy of the atoms, both in the condensate and
not belonging to it, can be expressed in terms of the characteristic
momentum as $p_{\min }^{2}/\left( 2m\right) $. The system may be considered
as effectively one-dimensional if this energy is much smaller than the
transverse excitation energy, $\omega _{\perp }$, i.e., $\omega _{\perp
}^{2}\gg 2\pi na_{\mathrm{bg}}\mu \Delta /m$, where the total initial
density of atoms, $n=m\omega _{\perp }\left( |\varphi _{0}\left( 0\right)
|^{2}+2|\varphi _{m}\left( 0\right) |^{2}\right) /\left( 2\pi \right) $, is
proportional to Manley-Rowe invariant (\ref{MR}), in terms of Eqs. (\ref{1}).

Further calculations involve a numerical solution of Eqs.\ (\ref{Psics}) on
a grid of values of $p$, combined with Eq.\ (\ref{phimf}). Figure \ref%
{FigBEC} presents the results for the 853 G Feshbach resonance in the
condensate of $^{23}$Na atoms, with $\Delta =0.01$ G, $a_{\mathrm{bg}}=3.4$
nm, and $\mu =3.65\mu _{B}$ (see Ref. \cite{YB03}), with $n=2\times 10^{11}$
cm$^{3}/$s and $\omega _{\perp }=1\times 2\pi $ KHz. The deactivation rate
coefficients, $k_{a}=5.5\times 10^{-11}$cm$^{3}/$s and $k_{m}=5.1\times
10^{-11}$ cm$^{3}/$s, were taken from Ref.\ \cite{MAXCK04}. The detuning $%
D_{1D}$ is defined as per Eq.\ (\ref{scaling}), with the harmonically
modulated mismatch parameter,
\begin{equation}
\alpha =\alpha _{0}+\Delta \alpha \cos \left( 2\pi z/L\right) ,
\label{harmonic}
\end{equation}%
cf. the modulation map in the optical model given by Eq. (\ref{2}). The
choice of the scaling factor in Eqs. (\ref{scaling}) which corresponds to
the Karamzin-Sukhorukov soliton, see Eqs. (\ref{Karamzin}), is $\Phi=\sqrt{%
8\pi n/\left( 27m\omega _{\perp }\right) }$. The value of $\alpha _{0}=0.992$
is chosen so as to make the initial molecular fraction equal to half the
total population, as at the center of the Karamzin-Sukhorukov soliton.

\begin{figure}[tbp]
\centering\subfigure[]{\includegraphics[width=3.375in]{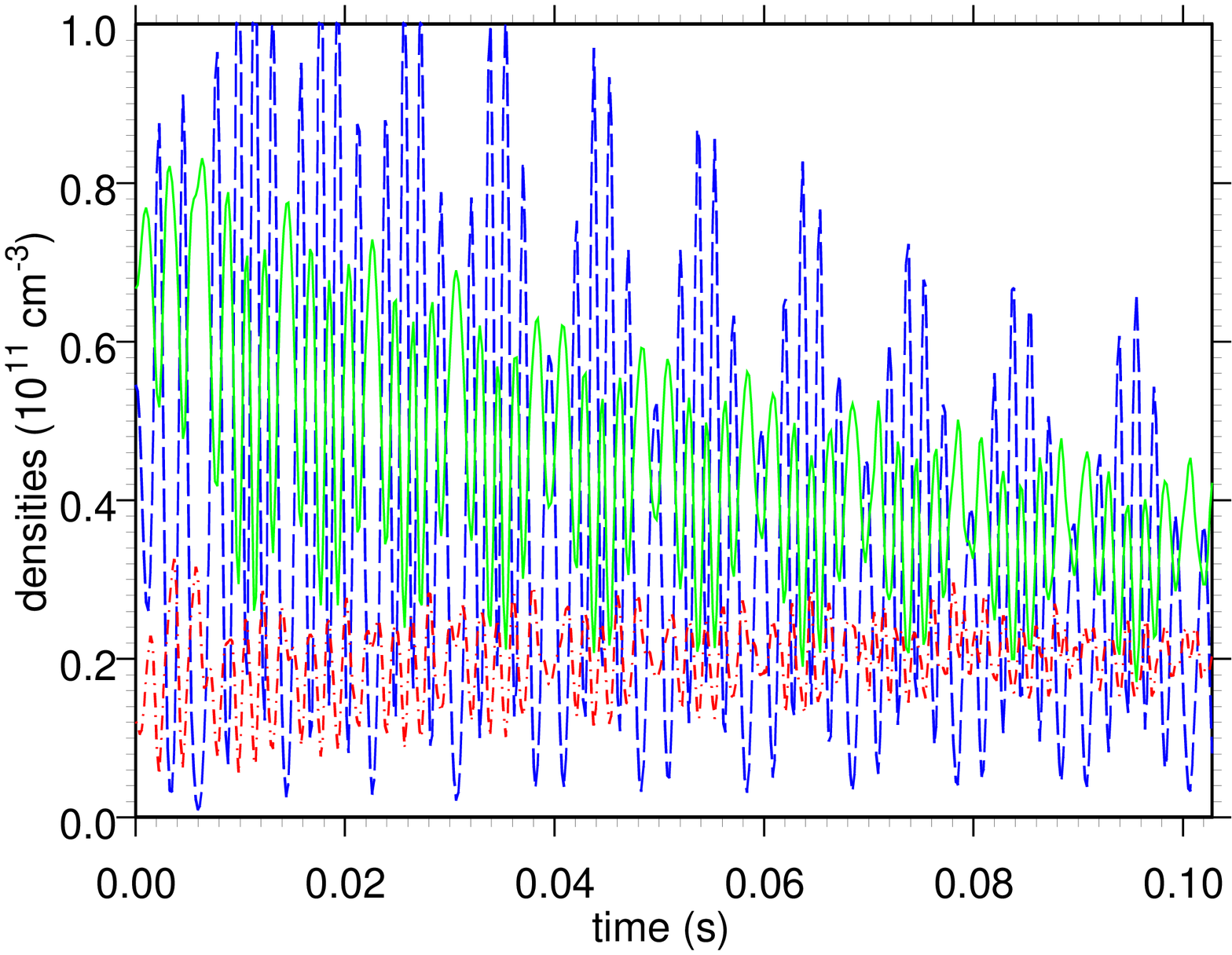}} \centering%
\subfigure[]{\includegraphics[width=3.375in]{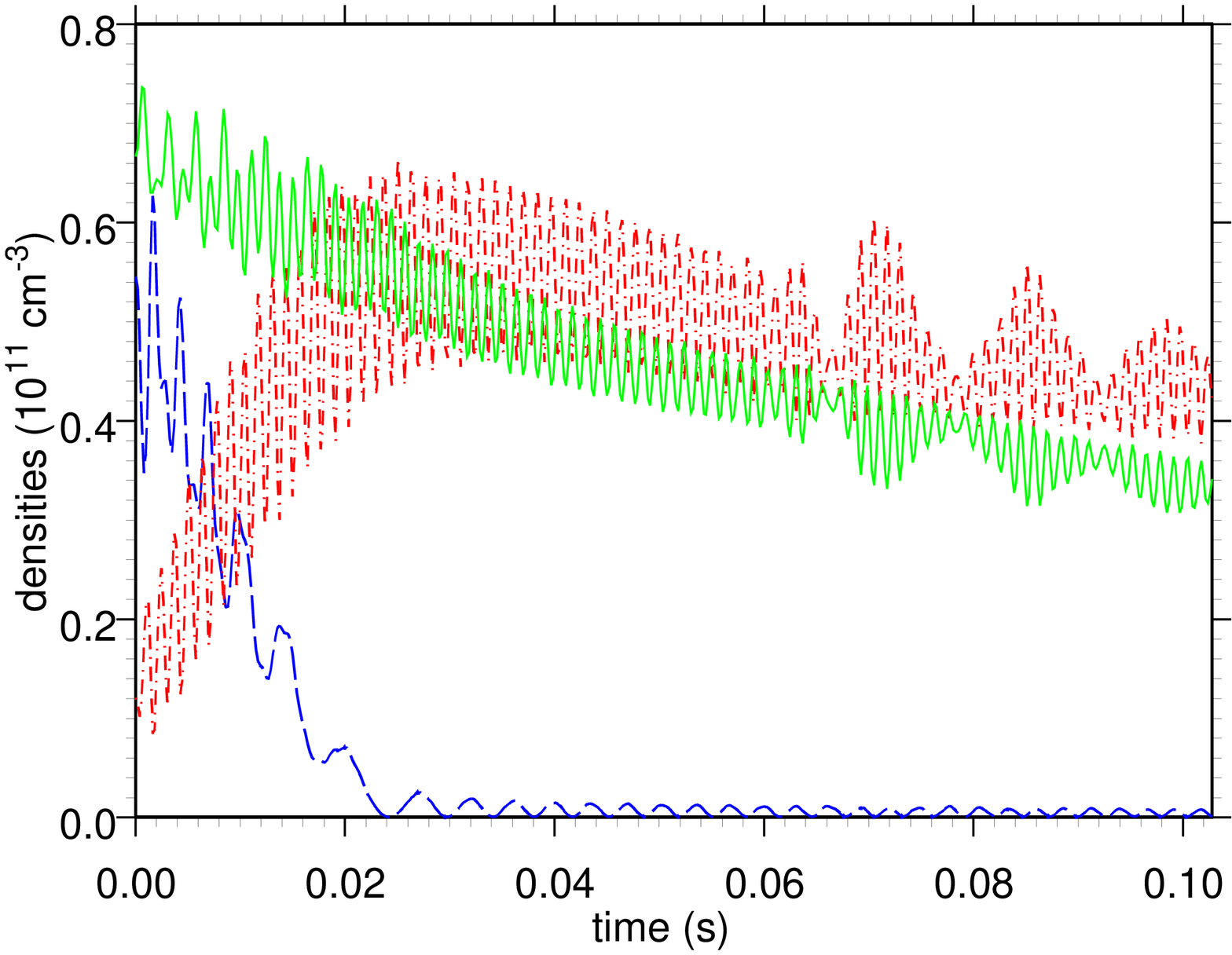}}
\caption{(Color online) Densities of the molecular and atomic
fractions in the condensate (solid and dashed lines), and of
non-condensate atoms (dot-dashed lines), calculated with $\Delta
\protect\alpha =1$ for $L=3$ (a) and $L=2$ (b), see Eq.
(\protect\ref{harmonic}).} \label{FigBEC}
\end{figure}

Figure \ref{FigBEC}(a) demonstrates that, for large modulation periods in
Eq. (\ref{harmonic}), the non-condensate fraction remains below the level of
$10\%$, and the lifetime due to the deactivation is large enough to allow
experimental observation of the soliton dynamics described in the previous
sections, in terms of the optical medium, in the hybrid atom-molecule BEC
too.

However, Fig.\ \ref{FigBEC}(b) demonstrates that the non-condensate fraction
of the atomic population acquires substantial gain for smaller modulation
periods. Therefore, the mean-field approach may not be applicable in this
region for the description of the atom-molecular quantum gas, and new
instability mechanisms can be expected.

\section{Conclusions}

In this work, we have proposed a model of the second-harmonic-generating ($%
\chi ^{(2)}$) system with the mismatch parameter subjected to the
periodic modulation (``mismatch management", MM). The system may
be implemented in two altogether different physical contexts: the
co-propagation of the FF (fundamental-frequency) and SH\
(second-harmonic) waves in a planar optical waveguide, and
atomic-molecular mixtures in the BEC (in the latter setting, the
atomic and molecular mean fields play the roles of the FF and SH
components, respectively). The most physically relevant approach
to the realization of the MM in these media are, respectively, a
long-period supermodulation imposed on top of the
quasi-phase-matching periodic arrangement of ferroelectric domains in the $%
\chi ^{(2)}$ optical waveguide, such as LiNbO$_{3}$, and the Feshbach
resonance tuned by a modulated magnetic field in the atomic-molecular BEC.
Accordingly, the natural form of the periodic modulation is
piecewise-constant in the former case, and harmonic in the latter one.

The main issue considered in the framework of the mean-field approach was
identification of the stability region for the mismatch-managed spatial
solitons in the plane of two control parameters, $\Delta \alpha $ and $L$
(the MM amplitude and period). In particular, a notable feature of the
stability area is the existence of an instability enclave embedded in it.
Also investigated was the robustness of the solitons against variation of
the shape of the input beam, and reduction of its power. It was found that
the stability of the established regime virtually does not depend on the
particular shape of the input, as well as on distribution of the total power
(Manley-Rowe invariant) between the FF and SH component in the input. On the
other hand, reduction of the initial peak power by a factor of $W<1$ reveals
the existence of a threshold, $W_{\mathrm{cr}}\approx 0.72$, below which the
initial pulse decays.

In the model of the atomic-molecular BEC, we have demonstrated that the
time-evolution equations for the mean fields are tantamount to the
spatial-evolution equations in the optical model. On the other hand,
important issues in the context of BEC are the stability of the condensate
against generation of fluctuational (non-condensate) components in the
degenerate quantum gas, and losses due to inelastic collisions. These
effects were analyzed within the parametric approximation, which goes beyond
the mean-field description. Numerical calculations have demonstrated that,
quite naturally, the condensate is effectively stable against the periodic
perturbations introduced by the MM in the low-frequency modulation format,
and unstable in the high-frequency regime.

\section*{Acknowledgement}

This work was supported, in a part, by the Israel Science Foundation through
a Center-of-Excellence grant No. 8006/03


\begin{thebibliography}{99}
\bibitem{KA} Yu. S. Kivshar and G. P. Agrawal, \textit{Optical Solitons:
From Fibers to Photonic Crystals} (Academic Press: San Diego, 2003).

\bibitem{ReviewRandy} K. E. Strecker, G. B. Partridge, A. G. Truscott, and
R. G. Hulet, 
New J. Phys. \textbf{5}, 73 (2003); F. Kh. Abdullaev, A. Gammal, A. M.
Kamchatnov, and L. Tomio, 
Int. J. Mod. Phys. B 19, 3415
(2005); O. Morsch and M. Oberthaler, Rev. Mod. Phys. \textbf{78}, 179 (2006).

\bibitem{DM} S. K. Turitsyn, E. G. Shapiro, S. B. Medvedev, M. P. Fedoruk,
and V. K. Mezentsev,
Compt. Rend. Phys. \textbf{4}, 145
(2003).

\bibitem{book} B. A. Malomed, \textit{Soliton Management in Periodic Systems}
(Springer:\ New York, 2006).

\bibitem{FRM} P. G. Kevrekidis, G. Theocharis, D. J. Frantzeskakis and B. A.
Malomed, Phys. Rev. Lett. \textbf{90}, 230401 (2003).

\bibitem{Fatkhulla} F. Kh. Abdullaev, J. G. Caputo, R. A. Kraenkel, and B.
A. Malomed, Phys. Rev. A \textbf{67}, 013605 (2003); H. Saito and M. Ueda,
Phys. Rev. Lett. \textbf{90}, 040403 (2003).

\bibitem{SSM} R. Driben and B.A. Malomed, Opt. Commun. \textbf{185}, 439
(2000); R. Driben, B. A. Malomed, and P. L. Chu, J. Opt. Soc. Am. B \textbf{%
219}, 143 (2003).

\bibitem{SSM2} R. Driben, B. A. Malomed, and P. L. Chu, Opt. Commun. \textbf{%
245}, 227 (2005).

\bibitem{WDM} R. Driben and B. A. Malomed, Opt. Commun. \textbf{197}, 481
(2001).

\bibitem{PMD} R. Driben and B. A. Malomed, ``Soliton stability against
polarization-mode-dispersion in the split-step system", Opt. Commun., in
press.

\bibitem{chi2} C. Etrich, F. Lederer, B. A. Malomed, T. Peschel, and U.
Peschel, 
in: Progress in Optics \textbf{41}, 483
(ed. by E. Wolf: North Holland, Amsterdam, 2000); A. V. Buryak, P. Di
Trapani, D. V. Skryabin, and S. Trillo,
Phys. Rep. \textbf{370}, 63 (2002).

\bibitem{Paolo} P. Di Trapani, D. Caironi, G. Valiulis, A. Dubietis, R.
Danielius, and A. Piskarskas, Phys. Rev. Lett. \textbf{81}, 570 (1998).

\bibitem{TTHK99} E. Timmermans, P. Tommasini, M. Hussein, and A. Kerman,
Phys. Rep. \textbf{315}, 199 (1999).

\bibitem{StoofReview} R. A. Duine and H. T. C. Stoof,
Phys. Rep. \textbf{396}, 115
(2004).

\bibitem{KGJ06} T. Koehler, K. Goral, P. S. Julienne, cond-mat/0601420; Rev.
Mod. Phys. (in press).

\bibitem{Peter} P. D. Drummond, K. V. Kheruntsyan, and H. He, Phys. Rev.
Lett. \textbf{81}, 3055 (1998).

\bibitem{Lluis} L. Torner,
IEEE Phot. Tech. Lett. \textbf{11}, 1268 (1999).

\bibitem{Dumitru} L. Torner, S. Carrasco, J. P. Torres, L. C. Crasovan, and
D. Mihalache, 
Opt. Commun. \textbf{199}, 277 (2001).

\bibitem{QPM} L. E. Myers, R. C. Eckardt, M. M. Fejer, R. L. Byer, W. R.
Bosenberg, and J. W. Pierce,
J. Opt. Soc. Am. B \textbf{12}, 2102-2116 (1995).

\bibitem{Warszawa} M. Trippenbach, M. Matuszewski, and B. A. Malomed,
Europhys. Lett. \textbf{70}, 8 (2005); M. Matuszewski, E. Infeld, B. A.
Malomed, and M. Trippenbach, Phys. Rev. Lett. \textbf{95}, 050403 (2005).

\bibitem{LH89} Y. Lai and H. A. Haus, Phys. Rev. A \textbf{40}, 854 (1989);
A. G. Shnirman, B. A. Malomed, and E. Ben-Jacob, Phys. Rev. A \textbf{50},
3453-3463 (1994).

\bibitem{YBO06} V. A. Yurovsky, A. Ben-Reuven, and M. Olshanii, Phys. Rev.
Lett. \textbf{96}, 163201 (2006).

\bibitem{YB03} V. A. Yurovsky and A. Ben-Reuven, Phys. Rev. A \textbf{67}
(2003).

\bibitem{HY00} H. A. Haus and C. X. Yu, J. Opt. Soc. Am. B \textbf{17}, 618
(2000).

\bibitem{walking} L. Torner, D. Mazilu, and D. Mihalache, Phys. Rev. Lett.
\textbf{77}, 2455 (1996); C. Etrich, U. Peschel, F. Lederer, and B. A.
Malomed, Phys. Rev. E \textbf{55}, 6155 (1997).

\bibitem{SKaramzin} Y. N. Karamzin and A. P. Sukhorukov, Pisma Zh. Exp.
Teor. Fiz. \textbf{20}, 730 (1974) [English translation: JETP Lett. \textbf{%
20}, 338 (1975)].

\bibitem{HS} H. Sakaguchi and B. A. Malomed, Phys. Rev. E \textbf{70},
066613 (2004).

\bibitem{Y05} V. A. Yurovsky, Phys. Rev. A \textbf{71}, 012709 (2005).

\bibitem{O98} M. Olshanii, Phys. Rev. Lett. \textbf{81,} 938 (1998).

\bibitem{MAXCK04} T. Mukaiyama, J. R. Abo-Shaeer, K. Xu, J. K. Chin, and W.
Ketterle, Phys. Rev. Lett. \textbf{92}, 180402 (2004).
\end{thebibliography}
\end{document}